\documentclass[aps,prb,twocolumn,psfig,showpacs]{revtex4}
\usepackage{amsfonts}
\usepackage{amsmath}
\usepackage{graphicx}
\usepackage{bm}
\usepackage{amssymb}
\usepackage{times}
\usepackage{dcolumn}
\usepackage{cases}
\usepackage{txfonts}
\usepackage{color}

\RequirePackage[hyperindex,colorlinks,bookmarksnumbered,plainpages=true]{hyperref}
\hypersetup{colorlinks,linkcolor=blue,urlcolor=blue,citecolor=blue}
\usepackage{hyperref}

\usepackage{ulem}
\renewcommand{\emph}[1]{\textit{#1}}

\definecolor{darkgreen}{rgb}{0,0.5,0}
\definecolor{darkblue}{rgb}{0,0,0.5}
\definecolor{darkred}{rgb}{.7,0,0}
\definecolor{purple}{rgb}{0.35,0,0.35}
\definecolor{orange}{rgb}{1,0.5,0}
\definecolor{grey}{rgb}{.6,.6,.6}

\newcommand{\Eq}[1]{Eq.~(\ref{#1})}

\newcommand{\Fig}[1]{Fig.~\ref{#1}}

\begin{document}

\title{Kosterlitz-Thouless transitions and phase diagrams of the interacting monomer-dimer model on a checkerboard lattice}
\author{Sazi Li$^{1}$, Wei Li$^{2,1}$, and Ziyu Chen$^{1,3,}$}
\email{chenzy@buaa.edu.cn}
\affiliation{$^{1}$ Department of Physics, Beihang University, Beijing 100191, China  \linebreak
$^{2}$ Physics Department, Arnold Sommerfeld Center for Theoretical Physics, and Center for NanoScience, Ludwig-Maximilians-Universit\"at, 80333 Munich, Germany
\linebreak
$^{3}$ Key Laboratory of Micro-nano Measurement-Manipulation and Physics (Ministry of Education), Beihang University, Beijing 100191, China}

\begin{abstract}
Using the tensor network approach, we investigate the monomer-dimer models on a checkerboard lattice, in which there are interactions (with strength $v$) between the parallel dimers on one-half of the plaquettes. For the fully-packed interacting dimer model, we observe a Kosterlitz-Thouless (KT) transition between the low-temperature symmetry breaking and the high-temperature critical phases; for the doped monomer-dimer case with finite chemical potential $\mu$, we also find an order-disorder phase transition which is of second-order, instead. We use the boundary matrix product state approach to detect the KT and second-order phase transitions, and obtain the phase diagrams $v-T$ and $\mu-T$. Moreover, for the non-interacting monomer-dimer model (setting $\mu=\nu=0$), we get an extraordinarily accurate determination of the free energy per site (negative of the monomer-dimer constant $h_2$) as $f=-0.662\, 798\, 972\, 833\, 746$ with the dimer density $n=0.638\, 123\, 109\, 228\, 547$, both of 15 correct digits.
\end{abstract}

\pacs{64.60.Cn, 05.50.+q, 05.10.Cc, 64.60.Fr}
\maketitle

\section{introduction}
Classical monomer-dimer model in two dimensions (2D) is one of the intriguing models in statistical mechanics. The problem has a venerable history, \cite{Fowler-1937, Fisher-1961, Kasteleyn-1961, Fisher-1963} it was firstly introduced in the context of the absorption of molecules on the surface.\cite{Fowler-1937} When a rigid molecule occupies two nearest neighbor (NN) sites on the (square) lattice, it can be regarded as a dimer linking the two NN sites, while an empty site means the presence of a monomer. The monomer-dimer model can be related to the Ising \cite{Fisher-1961} and the height models,\cite{height} playing an important role in the statistical physics. A special case of the monomer-dimer model, namely the fully-packed dimer model, can be analytically solved,\cite{Kasteleyn-1961, Fisher-1963, Baxter} while the general monomer-dimer case is not. Numerating all the possible configurations and calculating the properties of the monomer-dimer model is an NP-complete problem and thus unfortunately ``intractable" in computations.\cite{Jerrum, Huo, Baxter-1968, Kong, Friedland} Numerically, one has to adopt some approximate methods, like the Monte Carlo samplings, \cite{Alet-2005, Alet-2006} to study the monomer-dimer models.

The fully-packed dimer models exhibit different properties on bipartite and non-bipartite 2D planar lattices. The former supports a critical phase with the algebraic decaying dimer-dimer correlations, \cite{Fisher-1961,Kasteleyn-1961} while the latter (say, the triangular and kagome lattices) have exponential dimer-dimer correlations. \cite{Fendley, Krauth, Misguich} Exept the 2D lattices, people have also investigated the hard-core dimer models on various 3D lattices.\cite{Huse} Extended critical phases are found on the bipartite cubic lattice, while no critical phases are found on the non-bipartite 3D lattices.\cite{Huse} Moreover, the classical dimer models can be ``upgraded" to the so-called quantum dimer models, by promoting the classical dimer configurations to quantum state bases. The quantum dimer model was introduced by Rokhsar and Kivelson, \cite{RK-1988} where the singlet formed by two adjacent spins plays the role of a dimer. The quantum dimer model is one of the typical systems which exhibit nontrivial topology and fractional excitations. \cite{topological-1,topological-3,topological-4}

Recently, F. Alet \textit{et al.} introduce the interacting dimer models on a square lattice, where two parallel dimers on the same plaquette are coupled (attractive). They studied this interacting dimer model by Monte Carlo simulations, and found a dimer order-disorder phase transition, of Kosterlitz-Thouless (KT) type,\cite{KT} at a certain temperature.\cite{Alet-2005, Alet-2006} Interestingly, the introduction of interactions between dimers is not only of theoretical interest in the model study, but also acquire experimental realizations recently. An adsorption experiment of certain rodlike organic molecules on the graphite was reported \cite{experiment-2008} and it is found to be relevant to the fully-packed dimer model on a hexagonal lattice, with couplings between neighboring parallel dimers. \cite{experiment-2009}

\begin{figure}[htpb]
  \begin{center}
	\includegraphics[width=0.7\columnwidth]{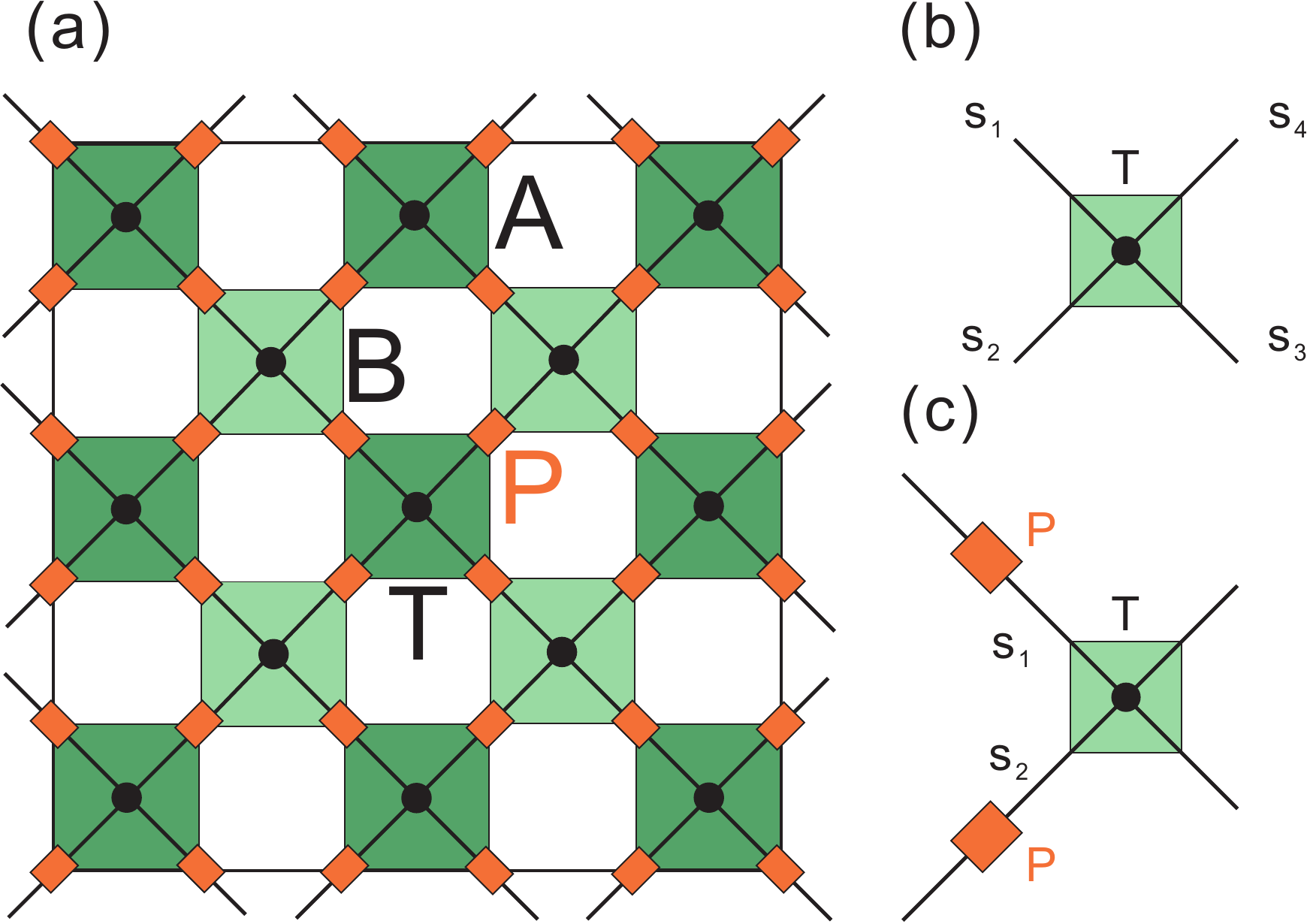}
  \end{center}
  \caption{(Color online) (a) The illustration of the monomer-dimer model on a checkerboard lattice. Interactions between the parallel dimers are introduced on the green (gray) plaquettes. The underlying tensor network is also shown, where the plaquette tensors ($T$) only cover one-half of the plaquettes. The square-lattice tensor network can be divided into two sublattices, namely A and B, denoted by dark and light green (gray) colors, respectively. Besides the plaquette tensor $T$, there is a local matrix $P$ on each vertex. (b) The plaquette tensor $T$ is shown, $s_i$ ($i \in \{1,2,3,4\}$) are the four indices of tensor $T$. (c) The vertex matrix $P$ can be absorbed into a plaquette tensor $T$ by contracting the sharing indices $s_1$ ($s_2$).}
  \label{fig-checkerboard}
\end{figure}

In this work, we introduce and study the interacting monomer-dimer model on a black-white checkerboard lattice. We find and employ the compact tensor network (TN) representation of the (grand) partition function to investigate the model. The problem of calculating the free energy is thus transformed into a problem of how to accurately contract the TN. In practice, we contract it row by row with the infinite time evolving block decimation (iTEBD) technique for the matrix product state (MPS) \cite{Vidal2007} and calculate the thermodynamic properties of the monomer-dimer model with high precision. Through numerical simulations, we show that the fully-packed dimer model has a low temperature dimer ordered phase and a high temperature critical phase, with a KT transition separating these two phases. We observe no singularity in the energy and its derivative (specific heat) curves, however, we detect the KT transition by calculating the order parameters and correlation length. On the other hand, when monomer doping is introduced in the model (with the chemical potential $\mu$), there is also an order-disorder phase transition at certain temperature, which is instead found to be of second order. Remarkably, in addition to the regular thermodynamic quantities including the energy derivatives and the order parameter, etc, we also detect the (second-order) phase transition by checking the ``entanglement" properties of the boundary MPS. We even extract the corresponding conformal central charge of the high-T critical phase of the fully-packed dimer model, by studying the block entanglement entropies of the boundary MPS. At last, collecting the phase transition points, we show and discuss the $\mu-T$ and $\nu-T$ phase diagrams of the interacting monomer-dimer model.

The rest of this paper is organized as follows. In Sec. II, we introduce the TN representation of the partition function and the method for  accurate evaluation of the thermodynamics. In Sec. III, we show the main numerical results on phase transitions, and the phase diagrams of the monomer-dimer model. The last section (Sec. IV) is devoted to a conclusion of the paper.

\section{Model and methods}
The interacting monomer-dimer model under study is defined on a black-white checkerboard lattice schematically depicted in \Fig{fig-checkerboard} (a), where the dimers are located on the links and occupy two lattice sites. Summing over all possible dimer coverings, we have the (grand) partition function as
\begin{equation}
\Xi = \sum_{\{c\}} \exp{[-\beta( v N_d- \mu N_{tot})]},
\label{eq-GPF}
\end{equation}
where $\{c\}$ means the set of all dimer configurations, $N_d$ counts the number of the doubly occupied green plaquettes (i.e., there are two parallel dimers on each plaquette). $\nu$ is the coupling strength between parallel dimers: $\nu<0$ is an attractive interaction, while $\nu>0$ means a repulsive one. In the follows, we only consider the attractive case, and set $\nu=-1$ as the energy scale if not otherwise specified. $N_{tot}$ is the total dimer number on the lattice, and $\mu$ is the chemical potential of dimers. Setting $\mu \to \infty$, we recover the fully-packed dimer model (no monomer doping).

\begin{figure}[htpb]
  \begin{center}
	\includegraphics[width=0.8\columnwidth]{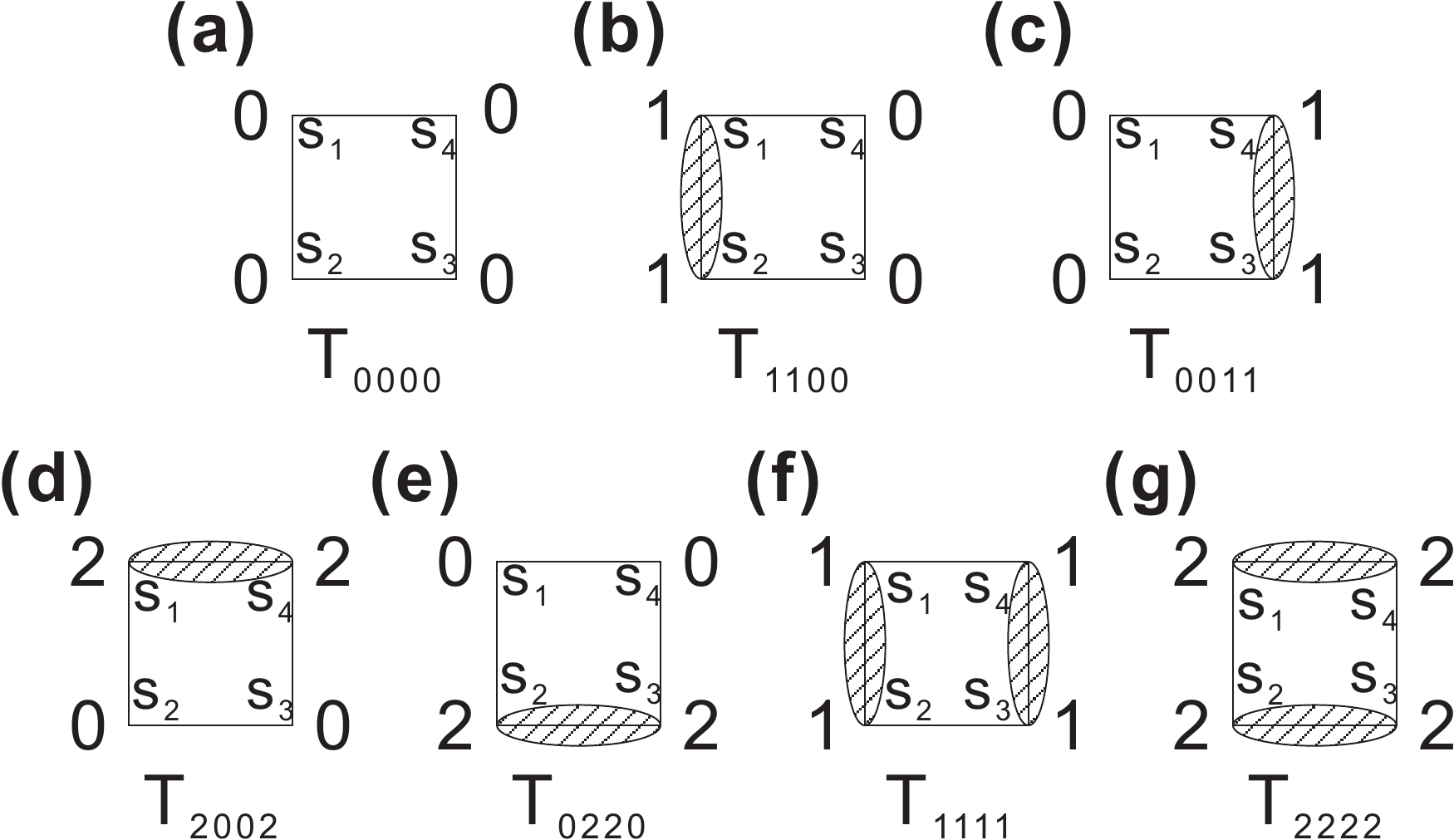}
  \end{center}
  \caption{The seven nonzero tensor elements of $T$ and the corresponding dimer configurations on a plaquette.
  (a) no dimer; (b-e) singlet dimer configurations; (f, g) double occupied plaquettes (vertical or horizontal).}
  \label{fig-config}
\end{figure}

The partition function of the monomer-dimer model has a simple tensor-network representation on a tilted square lattice. As shown in \Fig{fig-checkerboard}, the partition function TN consists of tensors $T_{s_1,s_2,s_3,s_4}$ located at one half of the plaquettes (the green ones) and the diagonal matrices $P_{s_i, s'_i}$ living on the vertices. $T$  has four bond indices $s_i$ ($i \in \{1,2,3,4\}$) corresponding to the vertices of the green plaquette. $s_i \in \{0,1,2\}$, meaning the presence ($s_i=1,2$) or the absence ($s_i=0$) of dimers on the concerned vertex $s_i$. In addition, we use $s_i=1,2$ to distinguish different dimers: $s_i=1$ means a vertical dimer, and $s_i=2$ is a horizontal one. Since the dimers are not allowed to touch each other (hard-core condition), there are only seven nonzero elements in the plaquette tensor $T$. The nonzero tensor elements $T$ and their corresponding classical dimer configurations are schematically shown in \Fig{fig-config} (a-g). $T_{0,0,0,0} = 1$ corresponds to the absence of any dimer on the plaquette, $T_{1,1,0,0} = T_{0,0,1,1} = T_{2,0,2,0} = T_{0,2,0,2} = 1$ represent the one-dimer configurations, and $T_{1,1,1,1} (T_{2,2,2,2}) = \exp(-\beta \nu)$ describes the plaquette with two vertical or horizontal dimers. The rest tensor elements are zero (thus forbidden to appear in the partition function).

In the fully-packed dimer case, to ensure that every vertex is occupied by one (and only one) dimer, a $3 \times 3$ matrix $P_{s_i,s'_i}$ is defined on every vertex, with elements $P_{0,1}=P_{1,0}=P_{0,2}=P_{2,0}=1$, otherwise 0. In the monomer-dimer case, we allow monomer doping in the model by setting $P_{0,0}=1$ and $P_{0,1}=P_{1,0}=P_{0,2}=P_{2,0} =\exp(\beta \mu/2)$, with $\mu$ the chemical potential. When $\mu \to \infty$, the monomer-dimer model recovers the fully-packed dimer case. Networking the plaquette tensors $T$ and vertex matrices $P$, we thus obtain a tensor network which faithfully represents the (grand) partition function of the interacting monomer-dimer model.

To calculate the thermodynamics, we adopt the infinite time evolution block decimation (iTEBD) method \cite{Vidal2007, Orus} for the accurate contraction of the partition function TN. iTEBD was proposed for efficient simulations of the time evolution and the ground state property (through imaginary-time evolution) of 1D quantum systems, and then generalized to calculate the thermodynamics of 2D classical statistical models \cite{Orus} and 1D quantum lattice models. \cite{LTRG} Within the boundary MPS framework, we utilize a kind of ``power" method to determine the dominating eigenvector (an MPS) of the transfer operator in the TN, which consists of a column of $T$ tensors organized in a matrix-product operator (MPO) form. However, unlike the ordinary power method for matrices, in the transfer MPO case the MPS is enlarged after each contraction step, with a composite bond space of a direct product of MPS and MPO bond bases. Thus the MPS bond dimension grows up exponentially with contraction steps. Therefore, one has to perform truncations on the bond space of the enlarged boundary MPS, and bring the bond dimension of MPS back to $D_c$, making the contraction procedure sustainable. Performing contraction and truncation processes iteratively until the boundary MPS converges, we thus obtain the dominating eigenvectors of the transfer MPO, with which we can then evaluate the expectation values of the local observables including the energy, the dimer occupation numbers, and the two-point correlation functions like dimer-dimer correlations.

In our practical calculations, we perform the contraction of MPS with transfer MPO until the prescribed convergence criterion is reached, say, free energy per site converges to $10^{-13}$ (in some cases even down to machine precision). The total number of iterations ranges between $4000$ and $10^{5}$, depending on the temperatures and the physical parameters of the model. The retained bond dimension of the boundary MPS $D_c \approx 100$ $\sim$ 150, the convergence with $D_c$ is always checked,  the truncation error is less than $10^{-6}$ at the critical point, and reaches the machine precision ($10^{-15}$) away from the critical points.

\begin{figure}
\centering
\includegraphics[width=0.45\textwidth]{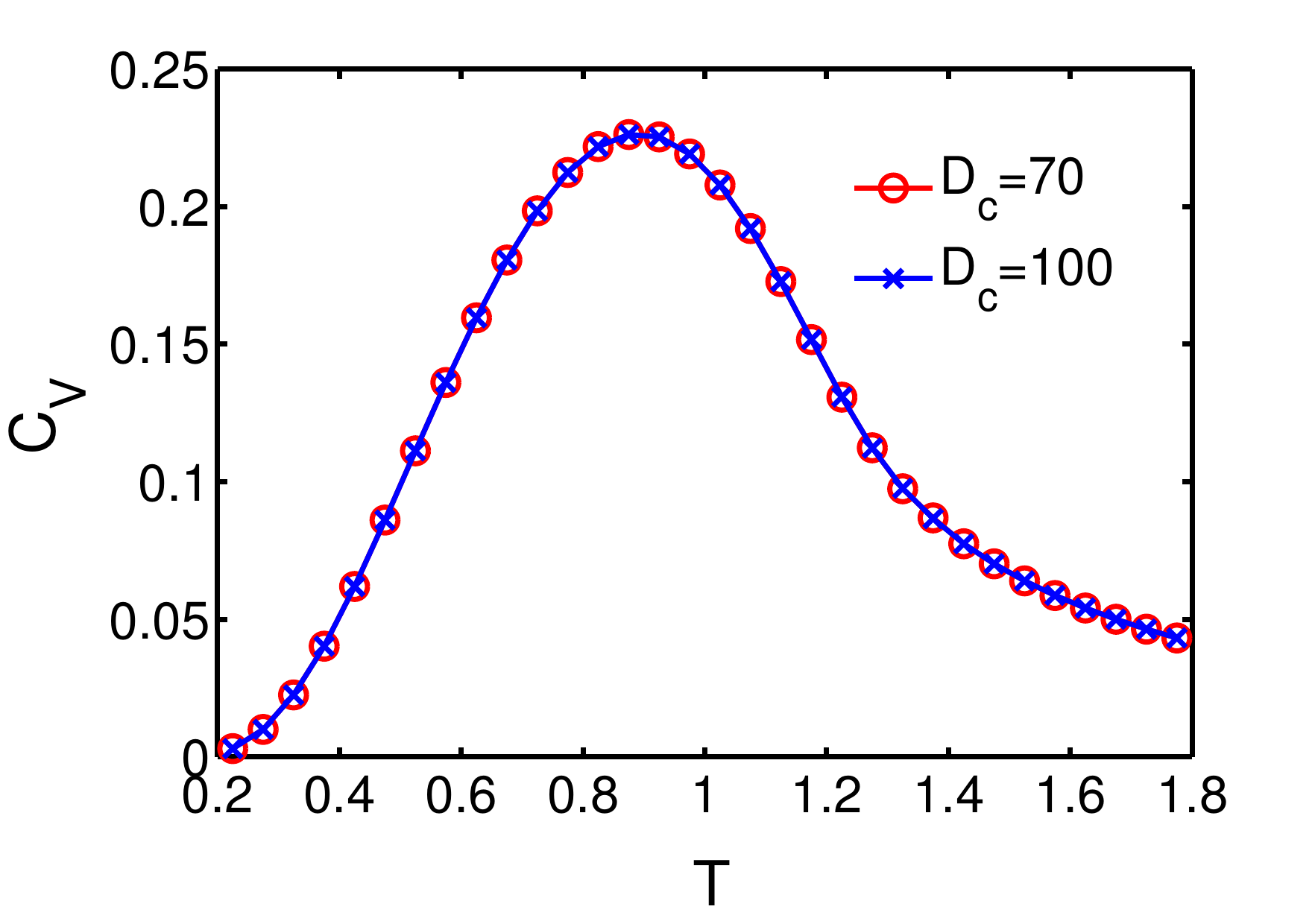}
\caption{(Color online) The specific heat $C_V$ of fully-packed dimer model with dimer-dimer interaction $\nu=-1$. The results are shown to be well converged with various retained bond states $D_c$.
\label{fully-Cv}}
\end{figure}

\begin{figure}
\centering
\includegraphics[width=0.4\textwidth]{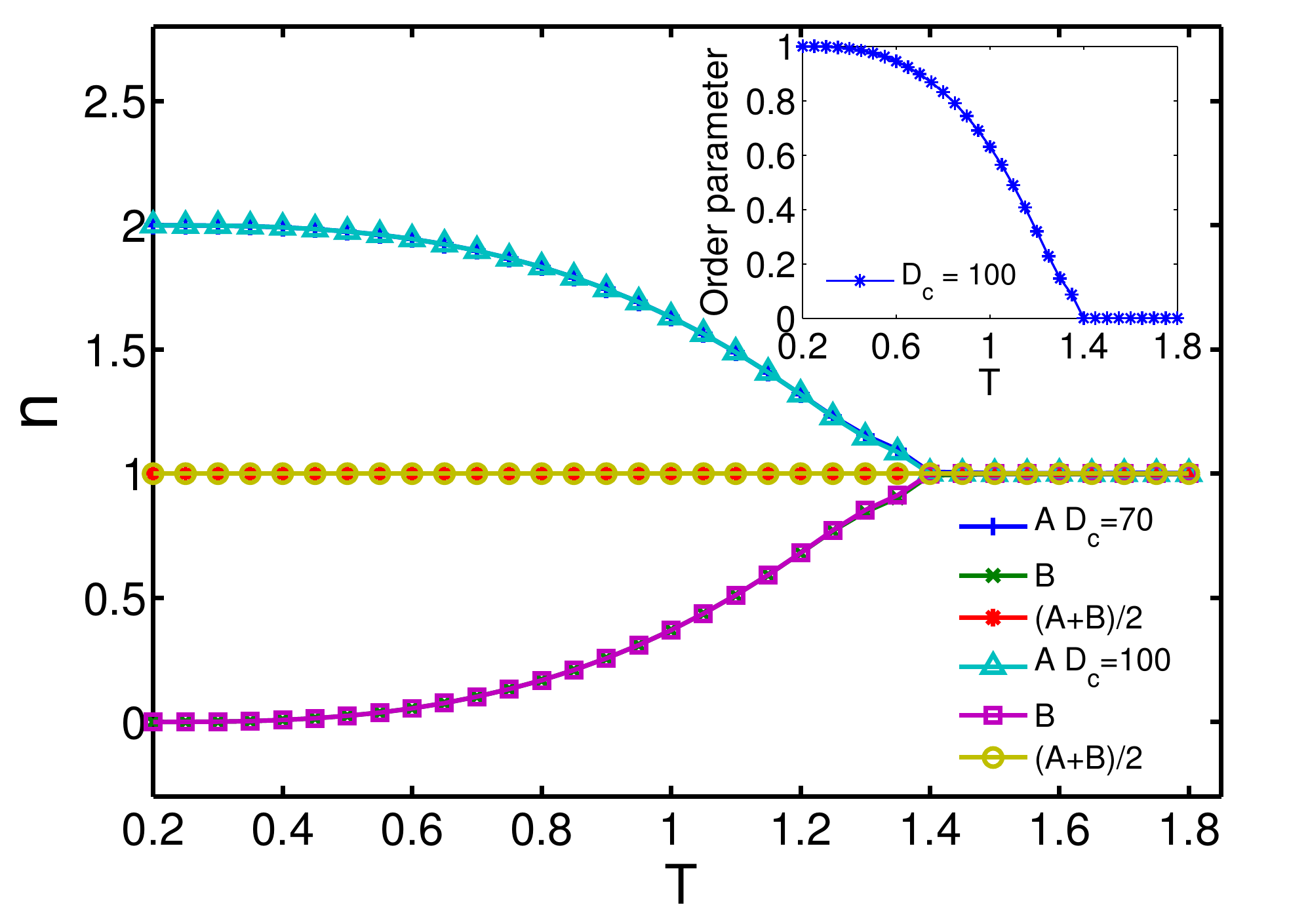}
\caption{(Color online) The expectation value $n_A$ ($n_B$) of the dimer occupation number on the green plaquettes A (B). The mean value $n = (n_A+n_B)/2$ is also plotted. The inset shows the order parameter $|n_A-n|$.
\label{fully-n}}
\end{figure}

\section{results and discussions}
\subsection{fully-packed dimer model}
Firstly, we investigate the interacting fully-packed dimer model on the checkerboard lattice. The specific heat $C_V$ curve is shown in \Fig{fully-Cv}, which is computed by taking first-order derivative (versus temperature $T$) of the energy per site. The latter is obtained by contracting the TN with one $T$ tensor replaced with an impurity tensor $T^I = \nu (T_{1,1,1,1} + T_{2,2,2,2}) $. From Fig. \ref{fully-Cv}, we observe no singularities in the $C_V$ curve, suggesting the absence of any second-order (or lower-order) phase transition.

However, by checking the local occupation number of the dimers on the green plaquettes (i.e., dimer density) in \Fig{fully-n}, we see different dimer densities $n_A \neq n_B$ between the A (dark) and B (light) green plaquettes [\Fig{fig-checkerboard} (a)] at low temperature $T < T_c \approx 1.4$. Especially in the limit $T\to0$, the A plaquette is filled with a pair of dimers ($n_A=2$) while the B plaquette is vacant ($n_B=0$). We use $n = (n_A+n_B)/2$ to denote the average dimer number on the green plaquettes. $n$ is verified to be a constant in the whole temperature region, this is because every site is linked to one dimer in the fully-packed case, and each green plaquette contains two sites in net. In the inset of \Fig{fully-n}, we show that the difference $|n_A - n|$ is nonzero below the critical temperature $T_c$, and it vanishes for $T>T_c$. Therefore, the particle number difference $|n_A - n|$ can be regarded as an order parameter detecting the phase transition between the low-T symmetry breaking phase ($n_A \neq n_B$) and a high-T disordered phase ($n_A$=$n_B$=1). Because the derivatives of energy are always continuous, this phase transition should belong to a KT type.\cite{KT}

\begin{figure}
\centering
\includegraphics[width=0.45\textwidth]{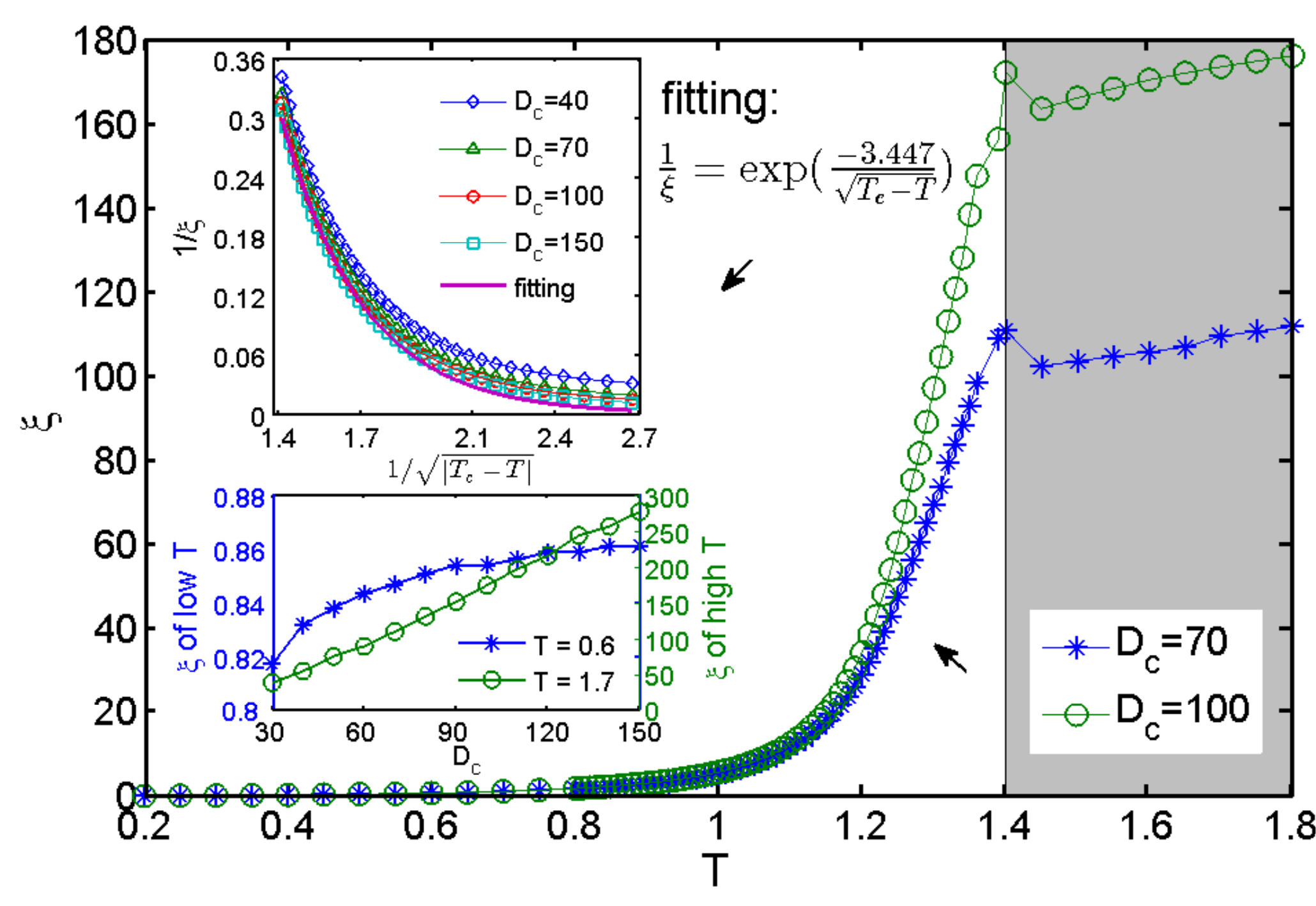}
\caption{(Color online) The correlation length $\xi$ versus temperature $T$. The correlations are measured between the green plaquettes of Fig. \ref{fig-checkerboard}, and the length unit is the distance between two nearest-neighboring plaquettes A (or B), i.e., twice the length unit of the original lattice. Upper inset shows $1/ \xi$ versus $1/ \sqrt{T_{c}-T}$ in the vicinity of $T_c$ ($T<T_{c}$), with the solid line an exponential fitting. The lower inset shows $\xi$ versus $D_c$ at $T=0.6<T_{c}$ and $T=1.7>T_{c}$, the former diverges while the latter saturates in the large $D_c$ limit.}
\label{fully-Xi}
\end{figure}

\begin{figure}
\centering
\includegraphics[width=0.42\textwidth]{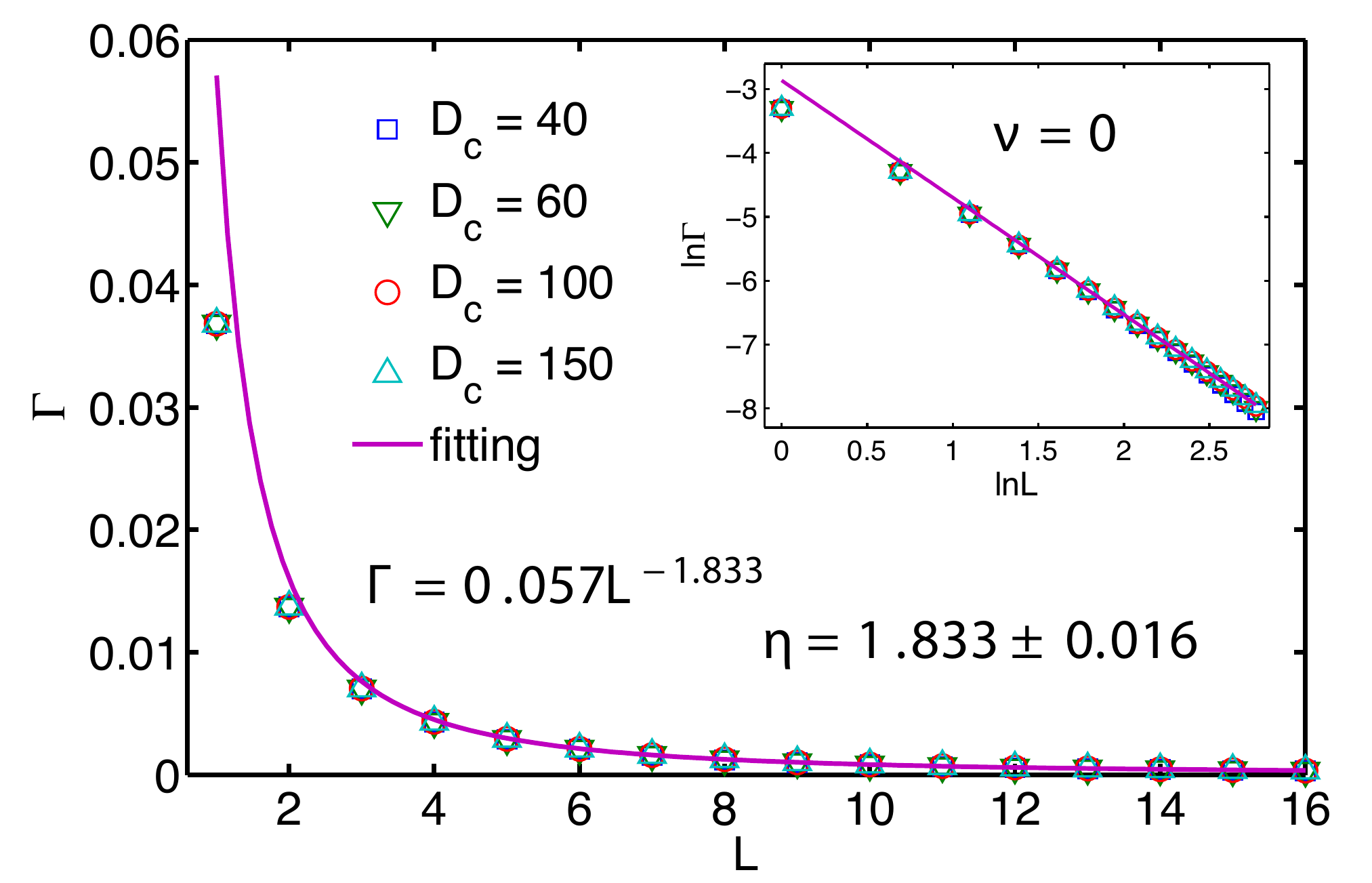}
\caption{(Color online) Log-log plot of the correlation function $\Gamma(L)$ with $\nu=0$ (or, equivalently $T=\infty$ limit) for the fully-packed dimer case. The fit $\Gamma(L) \sim L^{-\eta}$ reveals the algebraic decaying behavior (with $\eta$ obtained by linear fit shown in the inset).
\label{fully-v=0-func}}
\end{figure}

In order to understand this KT phase transition, we calculated the correlation length $\xi$ via the following formula,
\begin{equation}
\xi = 1/\ln (\frac{\lambda_{1}}{\lambda_{2}}),
\label{eq-xi}
\end{equation}
where $\lambda_{1}$ ($\lambda_{2}$) is the largest (second-largest) eigenvalue of the transfer matrix $M$, obtained by sandwiching two boundary MPS tensors
\begin{equation}
M_{a,b; a',b'} = \sum_m A_{a,b}^m (A^*)_{a',b'}^m,
\end{equation}
where $A$ is the MPS tensor, with $a,b$ the geometric indices and $m$ the sharing physical index. The results are shown in \Fig{fully-Xi}, where $\xi$ converges rapidly (with increasing $D_c$) to a finite value for $T<T_{c}$ (lower inset of \Fig{fully-Xi}), verifying the existence of a non-critical phase with finite correlation length. On the contrary, $\xi$ grows almost linearly with the increase of $D_c$ for $T\ge T_{c}$ (lower inset of \Fig{fully-Xi}). Therefore, we expect the $T\ge T_{c}$ region (shaded region in \Fig{fully-Xi}) is a critical phase with divergent correlation length. The upper inset of \Fig{fully-Xi} is the correlation length $\xi$ as a function of $1/ \sqrt{T_{c}-T}$ when $T$ approaches $T_c$ from below ($T<T_{c}$). It indicates that the correlation length $\xi$ diverges as an exponential of $1/ \sqrt{|T-T_{c}|}$.

\begin{figure}
\centering
\includegraphics[width=0.5\textwidth]{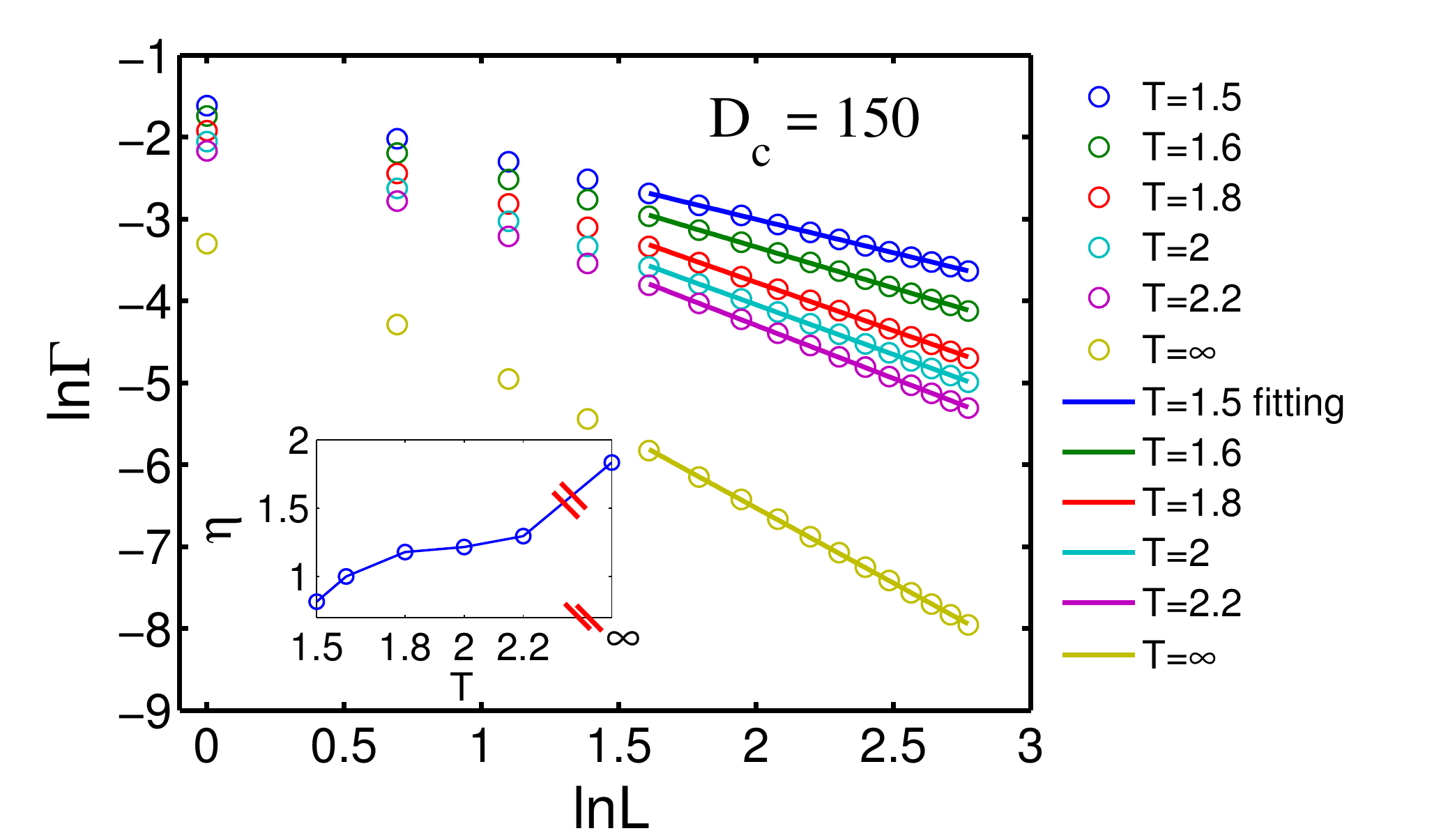}
\caption{(Color online) The correlation function at different temperatures $T$ ($>T_c$) for the fully-packed dimer model. Inset: the critical exponent $\eta$ versus $T$, $\eta$ is extracted from the fit ($\Gamma \sim L^{-\eta}$).}
\label{fully-hT-func}
\end{figure}

In Figs.~\ref{fully-v=0-func} and \ref{fully-hT-func}, we present explicitly the dimer number correlation function
\begin{equation}
\Gamma_{i,j} = \langle O_i O_j \rangle,
\label{eq-cf}
\end{equation}
where $O_i = (n_A)_i-  n$. In \Fig{fully-v=0-func}, we set $\nu=0$, and the partition function is an equal weight superposition of all possible fully-packed dimer configurations (corresponding to the $T=\infty$ limit). A log-log plot of $\Gamma$ versus $L$ and its algebraic fit ($\Gamma(x) \sim x^{-\eta}$) are shown in \Fig{fully-v=0-func}, where the algebraic decay is clearly verified. By  linear fitting, the exponent $\eta$ can be obtained, and is shown in \Fig{fully-v=0-func}. The deviations of $\Gamma(L)$ from the algebraic behavior (due to numerical errors) can be continuously corrected by increasing $D_c$. Notice that in Ref. \onlinecite{Alet-2006}, a related exponent is determined to be $\eta=2$ for both longitudinal and odd transverse dimer-dimer correlations (and $\eta=4$ for even transverse correlations), from analytical results. Our value $\eta\simeq1.83$ is obtained from dimer occupation number correlations Eq.~\ref{eq-cf} (horizontal and vertical dimers are not distinguished), which nevertheless well agrees with theirs.

Besides the $T= \infty$ ($\nu=0$) limit, we also studied correlations of other points in the critical phase with nonzero $\nu$, and present the results in \Fig{fully-hT-func}, from which we can observe algebraic decaying behaviors of $\Gamma$ for every $T > T_c$. In the inset of \Fig{fully-hT-func}, we show that the critical exponent $\eta$ grows monotonously with increasing $T$.

Therefore, by studying the dimer-dimer correlation function and correlation length, we find the phase transition occurring at $T_c$ is between an ordered (dimer crystal) phase and a critical (algebraic liquid) phase. This scenario also perfectly supports the conclusion of a KT-transition at $T_c$.

To gain further insight into the underlying physics of this critical phase in the fully-packed dimer model, we also extract the conformal central charge of this system in the critical regime $T>T_c$. The conformal field theory (CFT) tells us the conformal invariance at the critical point, and sets useful constraints on the critical behaviors of two-dimensional classical or 1D quantum systems \cite{CFT}. The universality class can be characterized by the conformal anomaly or central charge $c$ of the Virasoro algebra. We use MPS-based method to calculate the central charge, by fitting the block entanglement entropy $S$ versus the block size $L$. Depending on whether the system is in a critical or a noncritical regime, the block entanglement entropy has different asymptotic behaviors \cite{entanglement-1,entanglement-2}. In noncritical regimes, $S$ grows monotonously with $L$ before saturation; while in critical regimes, the CFT predicts a logarithmic divergence \cite{CFT-critical}
\begin{equation}
S \approx \frac{c}{3} \log_{2}(L)+k,
\label{eq-SL}
\end{equation}
where $L$ measures the site number of the block embedding in an infinite MPS, $c$ is the central charge and $k$ is a non-universal constant.

\begin{figure}
\centering
\includegraphics[width=0.45\textwidth]{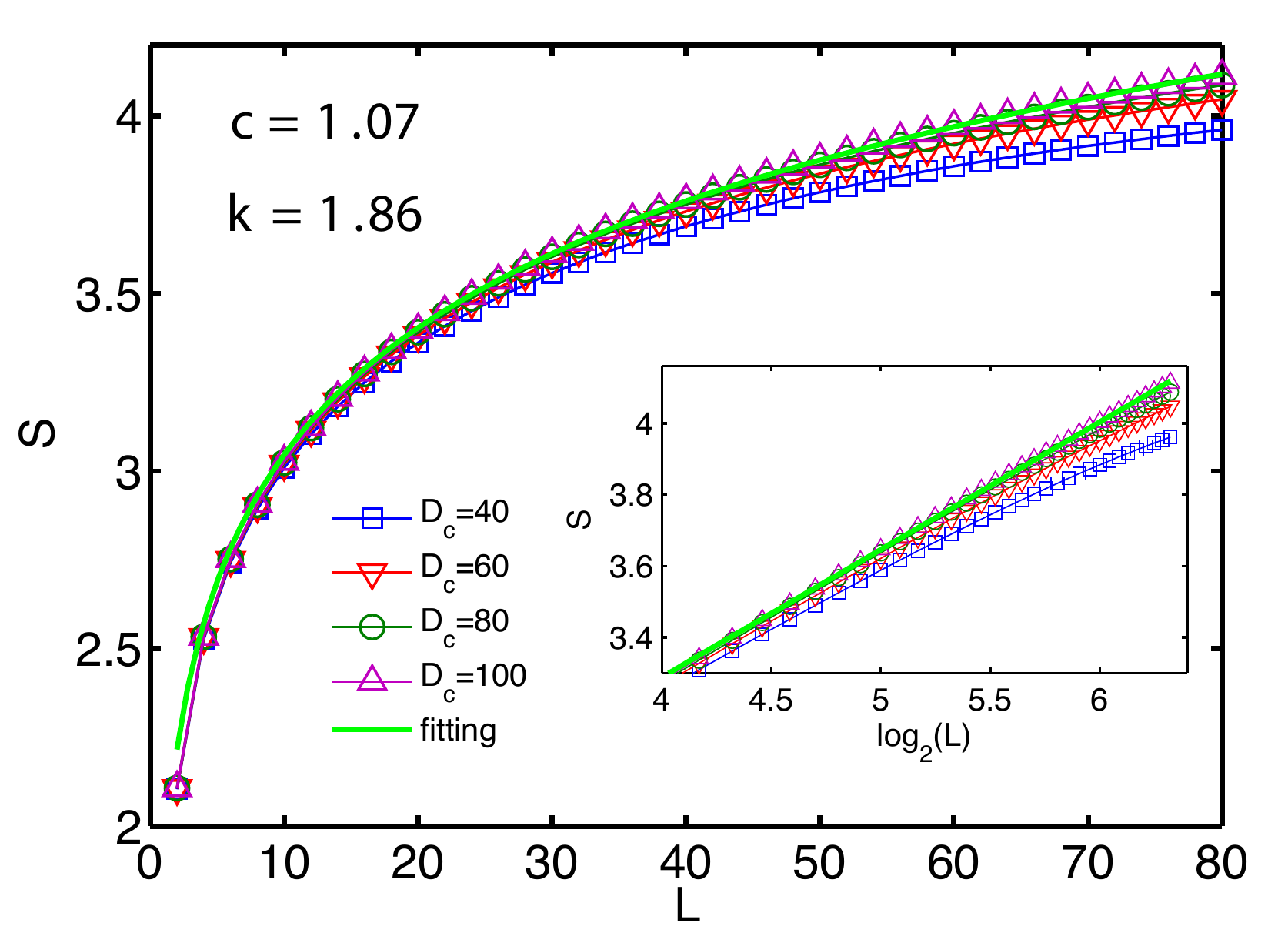}
\caption{(Color online) The block entanglement entropy $S$ as a function of length $L$ calculated by the iTEBD for the fully-packed dimer model, with $D_c = 40, 60, 80, 100$. The solid line is the fitting curve to \Eq{eq-SL} (obtained by linear fit shown in the inset), giving the result of central charge $c \simeq 1$. The inset: the entanglement entropy $S$ versus $\log_{2}(L)$.}
\label{charge}
\end{figure}

The entanglement entropy $S$ is defined by
\begin{equation}
S = - \rm{Tr}(\rho \log_{2} \rho),
\label{eq-rho}
\end{equation}
where $\rho$ is the reduced density matrix (DM) of system and can be calculated from the converged MPS. However, notice that for any $L$, the dimension of the reduced DM supported by the MPS is $D_c^2 \times D_c^2$. Therefore it is not possible to capture the entanglement entropies for extremely long $L$; however, by increasing the $D_c$ we are able to simulate the logarithmic divergence for sufficiently long $L$. By fitting our numerical results to \Eq{eq-SL}, as shown in \Fig{charge}, we find $c = 1$ for the fully-packed dimer model, in accordance with the result obtained by another independent method in Ref. \onlinecite{Alet-2005, Alet-2006}.

\subsection{monomer-dimer model}

\begin{figure}
\centering
\includegraphics[width=0.44\textwidth]{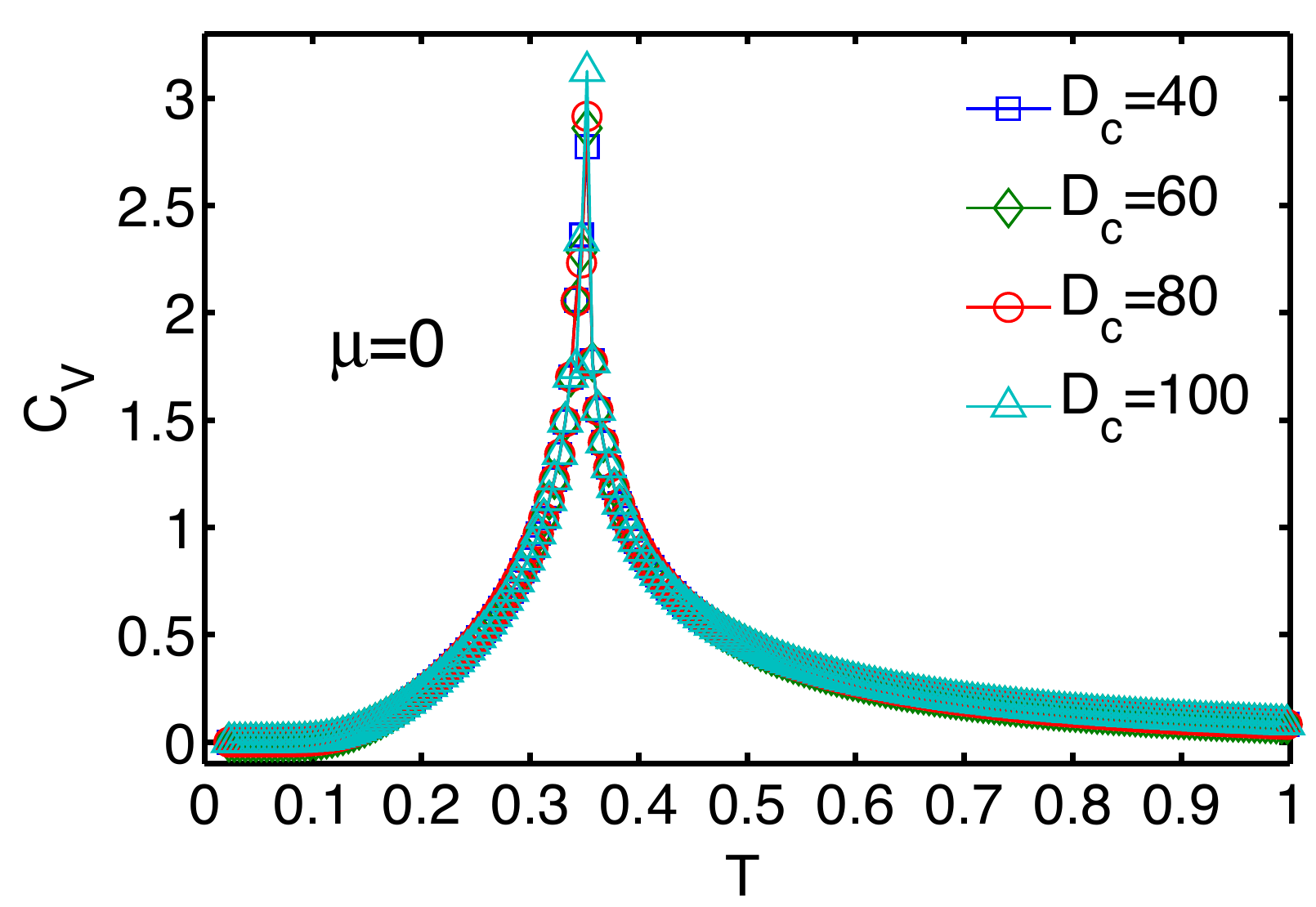}
\caption{(Color online) The specific heat $C_V$ of  the monomer-dimer model with a chemical potential $\mu=0$.
\label{u-Cv}}
\end{figure}

In this part, we study the interacting monomer-dimer model (with $\mu < \infty$) on the checkerboard lattice. \Fig{u-Cv} shows the calculated specific heat $C_V$ of the case $\mu=0$, where a divergent peak of $C_V$ occurs at $T_c=0.35$, uncovering the existence of a second-order phase transition.

\begin{figure}
\centering
\includegraphics[width=0.42\textwidth]{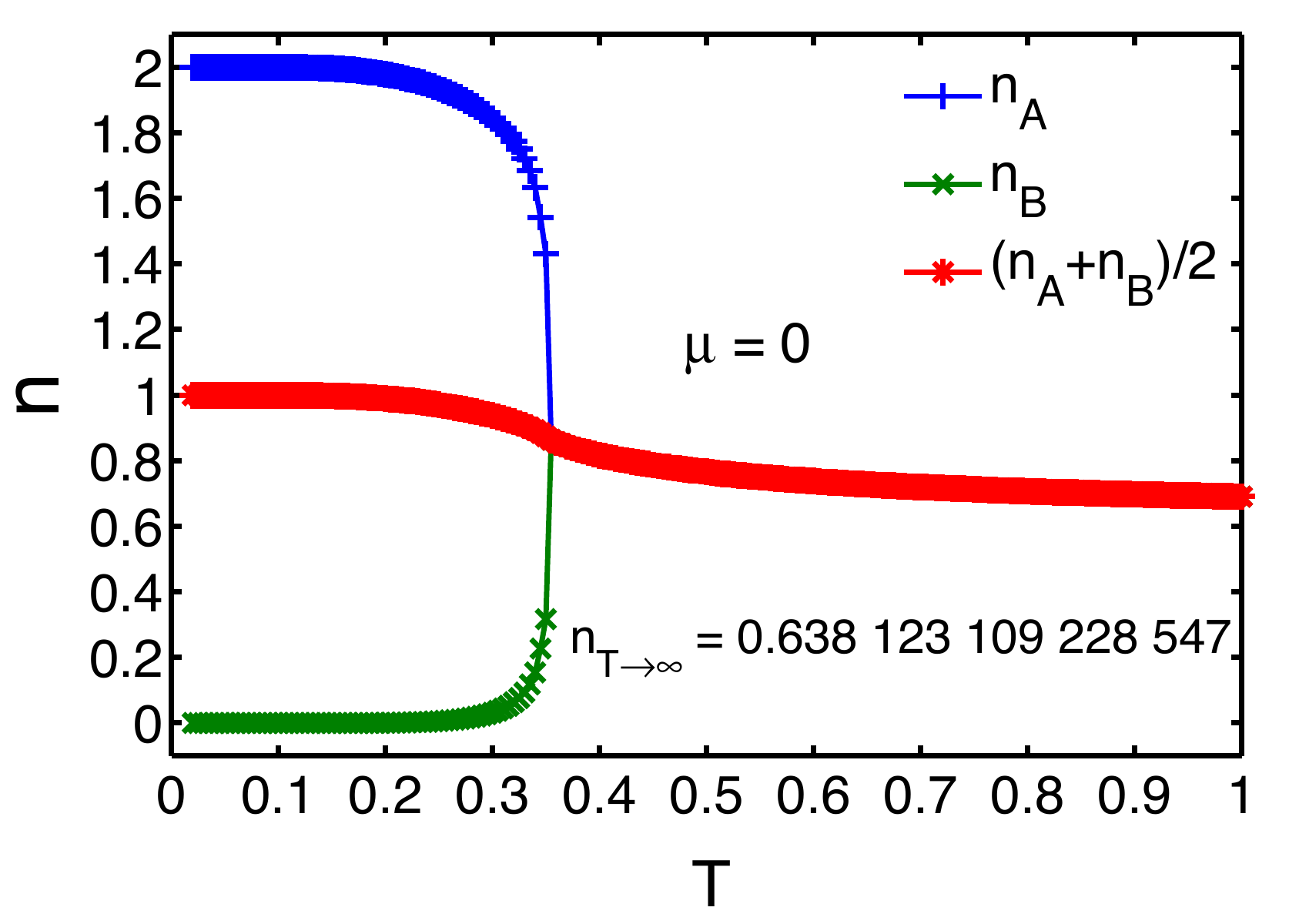}
\caption{(Color online) The dimer occupation number on the plaquettes A(B) and the average value $n$ for  $\mu=0, \nu=-1$.
\label{u-n}}
\end{figure}

The dimer occupation numbers $n_A$ ($n_B$) on the plaquettes A (B) and the average $n=(n_A+n_B)/2$  are shown in \Fig{u-n}. For $T<T_c$, the symmetry between the A and B plaquettes is broken ($n_A \neq n_B$), while for $T\ge T_c$ this symmetry is recovered ($n_A = n_B$). In contrast to the fully-packed case ($n=1$ as a constant), $n$ decreases with increasing temperatures in the monomer-dimer case. The limit $T \to \infty$ (or, equivalently $\mu=\nu=0$) is an interesting special case, i.e., the conventional (noninteracting) monomer-dimer model. The mean value $n$ is determined as $n_{T\to \infty} = 0.638\, 123\, 109 \, 228\,547$, which agrees perfectly with the previous studies (0.638 12311 in Ref. \onlinecite{Baxter-1968}, and 0.638 1231 in Ref. \onlinecite{Kong}), and provides 15 very well converged (correct) digits. The corresponding free energy per site (negative of the monomer-dimer constant $h_2$) is $f = -0.662\,798\,972\,833\,746$ with 15 converged (correct) digits, again in perfect agreements with previous results ($h_2=$-0.662 798 972 834 in Ref. \onlinecite{Kong}, and $h_2=$0.662 798 972 7 $\pm$ 0.000 000 000 1 in Ref. \onlinecite{Friedland}).

\begin{figure}
\centering
\includegraphics[width=0.45\textwidth]{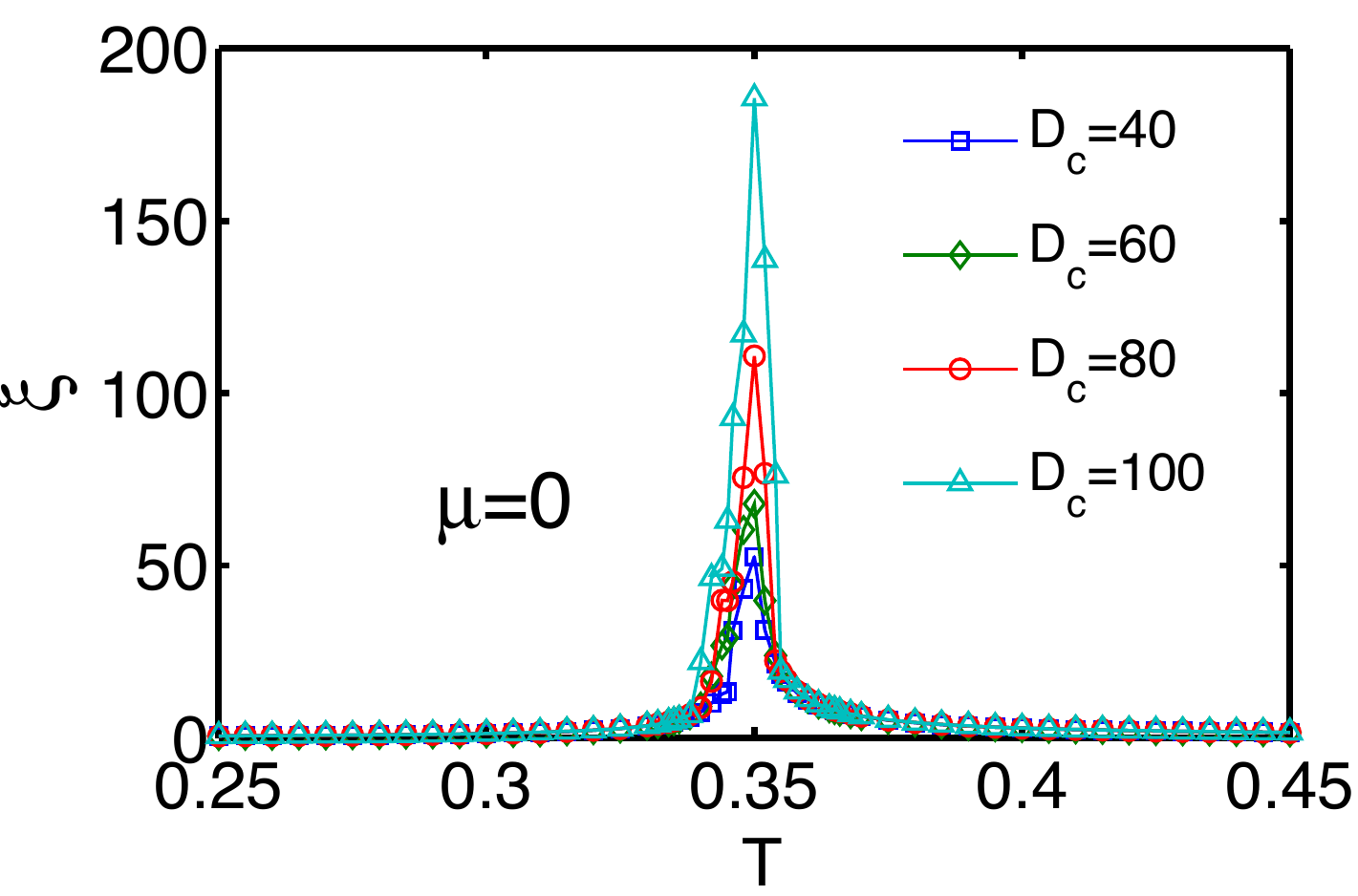}
\caption{(Color online) The correlation length for $\mu=0, \nu=-1$. The heights of the peaks at $T_c$ grow with the increase of $D_c$.
\label{u-Xi}}
\end{figure}

In \Fig{u-Xi}, we show the correlation length $\xi$, which also shows a divergent peak at $T_{c}$, the second-order phase transition point. Notice that in the $T>T_c$ region, the correlation length $\xi$ is finite, in contrast to the fully-packed case where $\xi$ diverges in the high-T disordered phase.

\begin{figure}
\centering
\includegraphics[width=0.45\textwidth]{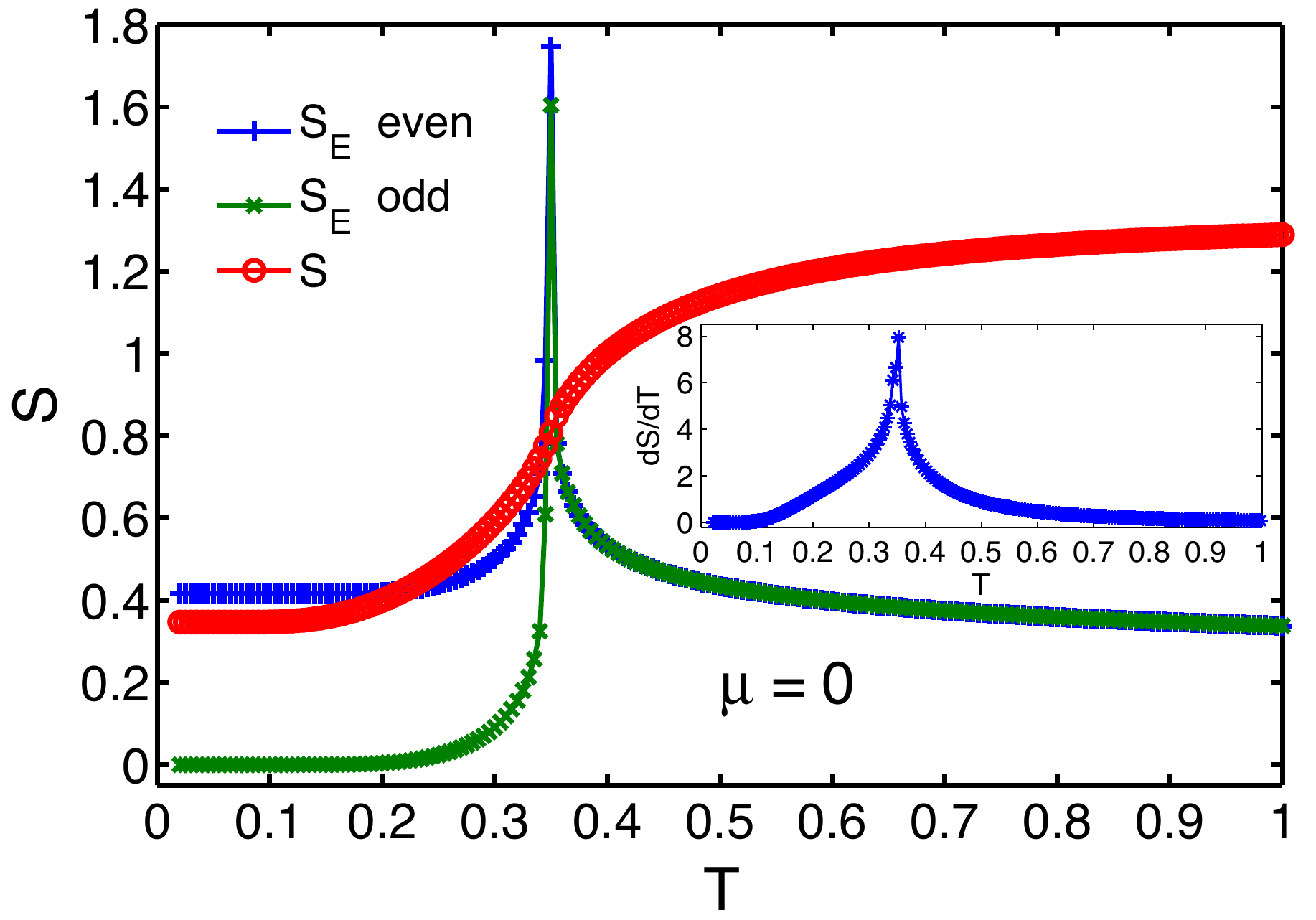}
\caption{(Color online) The entropy $S$ and the entanglement entropy $S_E$ for $\mu=0$. Because there are A and B plaquettes, the boundary MPS is of period two, leading to even and odd cut entropies $S_E$ even and odd. Inset: the differential of entropy $S$ for temperature $T$ $dS/dT$.
\label{u-S}}
\end{figure}

Entropy is another interesting quantity. Actually, we refer to two kinds of entropies in the calculations, i.e., the conventional thermodynamic entropy $S = (U-F)/T$ and the formal ``entanglement entropy" $S_E$ evaluated from the boundary MPS.  Given the boundary MPS, we can take a Schmidt decomposition (once for all bonds) of the translation-invariant MPS and formally calculate its ``entanglement properties". Notice that this bipartite entanglement entropy is different the block entanglement entropy discussed above, because the former is between two half-infinite chain. As shown in \Fig{u-S}, the bipartite entanglement entropy $S_E$ shows a clearly divergent peak at $T_c$, indicating the occurrence of a phase transition. This observation is quite remarkable, because the conventional thermodynamic entropy $S$ is smooth around $T_c$, and its singularity can only be seen after taking a derivative over $T$ (inset of \Fig{u-S}), owing to $\frac{\partial S}{\partial T} = \frac{C_{V}}{T}$. Therefore, this ``entanglement entropy" $S_E$ is found to be more sensitive to the phase transition (than the thermodynamic entropy $S$), and thus can serve as an useful numerical tool detecting continuous phase transitions.

\begin{figure}
\centering
\includegraphics[width=0.5\textwidth]{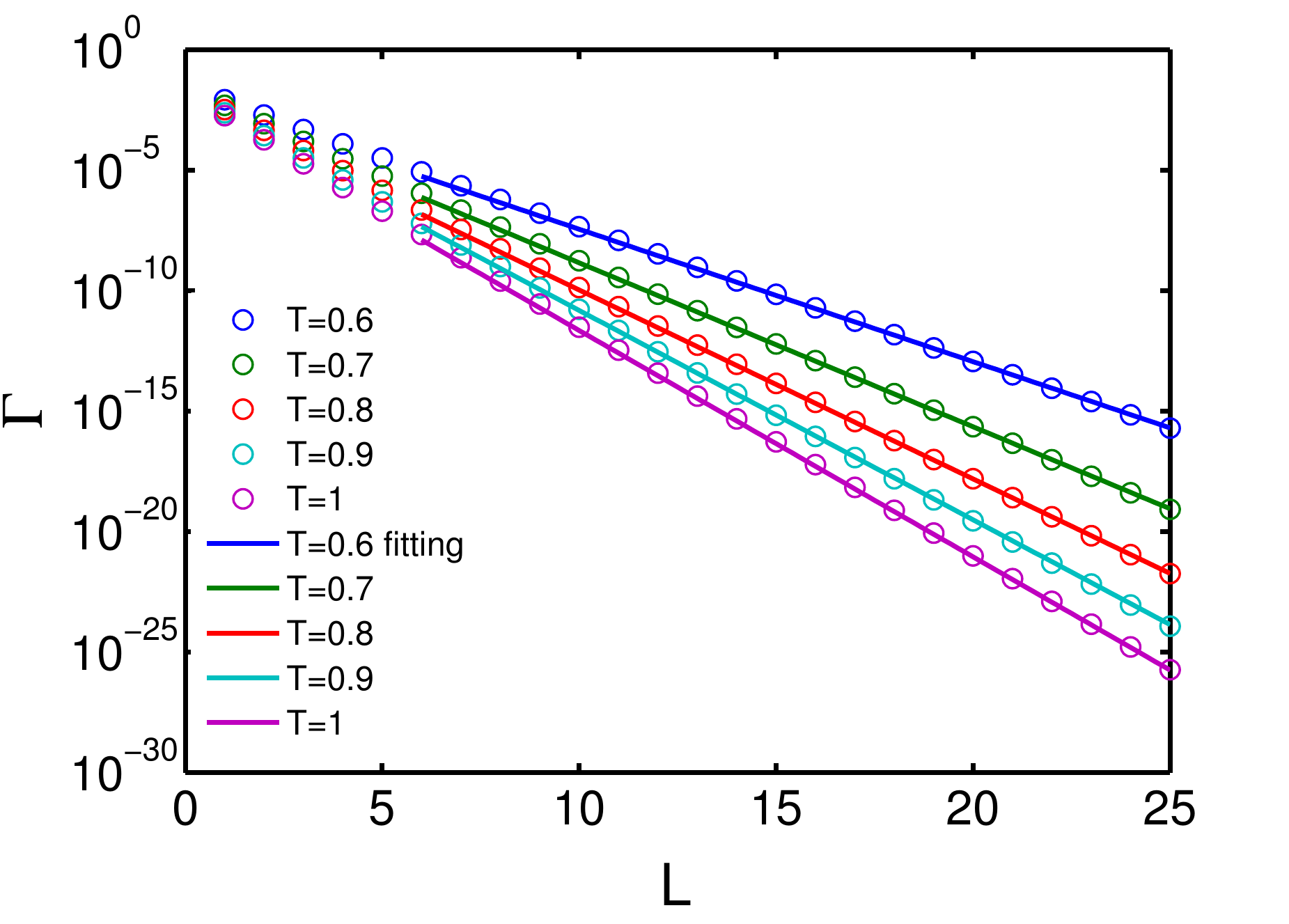}
\caption{(Color online) The correlation function $\Gamma$ for different temperatures higher than $T_{c}$ at $\mu=0$ and their fitting. The fitting used correlation length $\xi$ calculated directly (see details in the text). The fitting formula is $\Gamma(L) \sim e^{-L/\xi}$.
\label{u=0-func}}
\end{figure}

In \Fig{u=0-func}, we show the semi-log plot of the correlation functions $\Gamma$ for $\mu=0$. The linear fittings are performed using the correlation length $\xi$ estimated from the transfer matrices (\Eq{eq-xi}). Note that for both $T>T_c$ and $<T_c$ cases, the correlation functions are exponentially decaying, indicating that the high-T phase is non-critical under the monomer doping.

\begin{figure}
\centering
\includegraphics[width=0.5\textwidth]{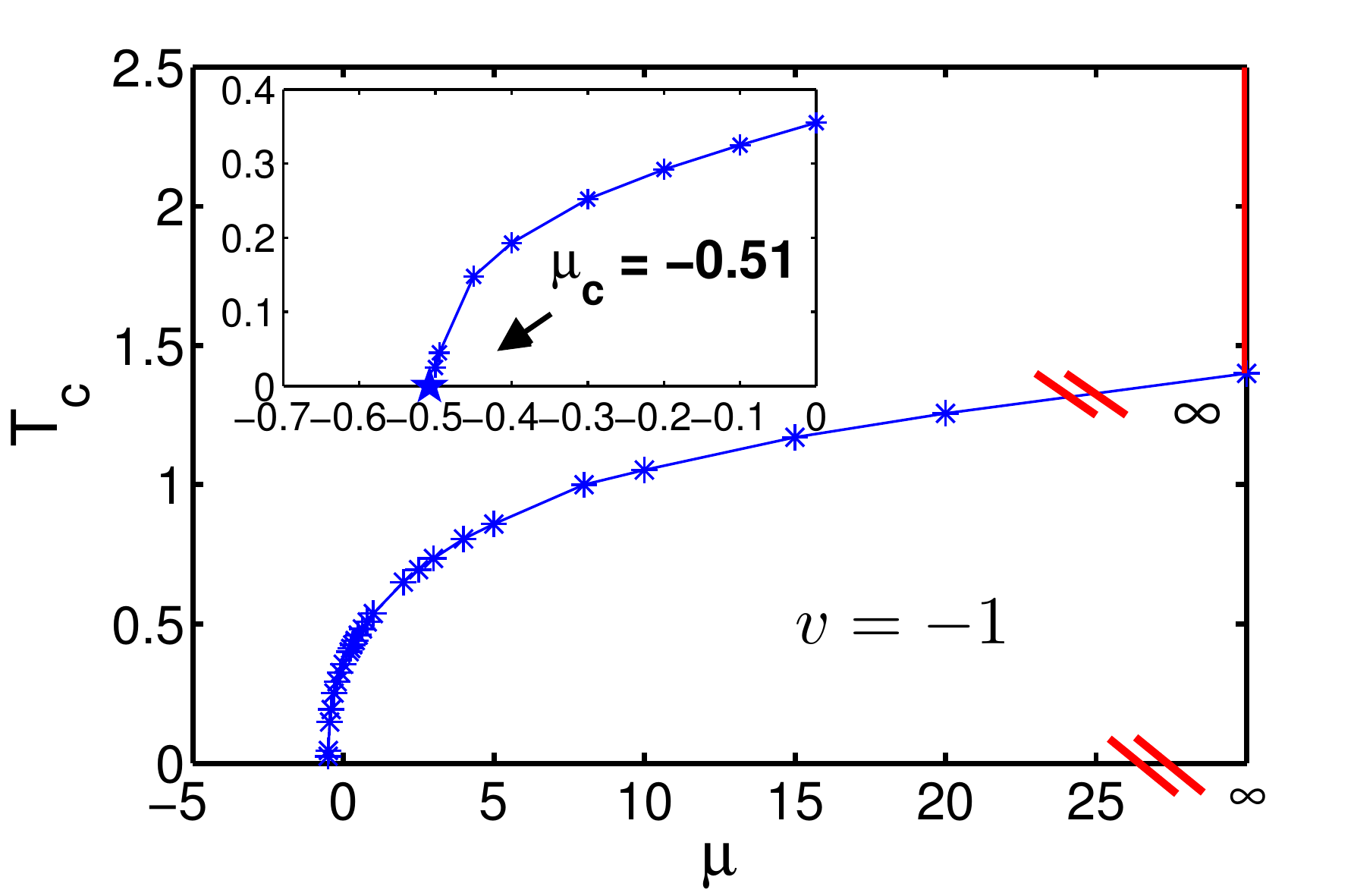}
\caption{(Color online) The phase diagram of the monomer-dimer model on the checkerboard lattice with $v=-1$. The (red) vertical line ($\mu = \infty$) is a critical line; while the blue curved line (with finite $\mu$) is a second order phase transition line. Inset: Amplification of the negative $\mu$ region, the star denotes the terminating point $\mu_{c} \approx -0.51$ of the critical line.}
\label{phase-u}
\end{figure}

\subsection{phase diagram}

\begin{figure}
\centering
\includegraphics[width=0.5\textwidth]{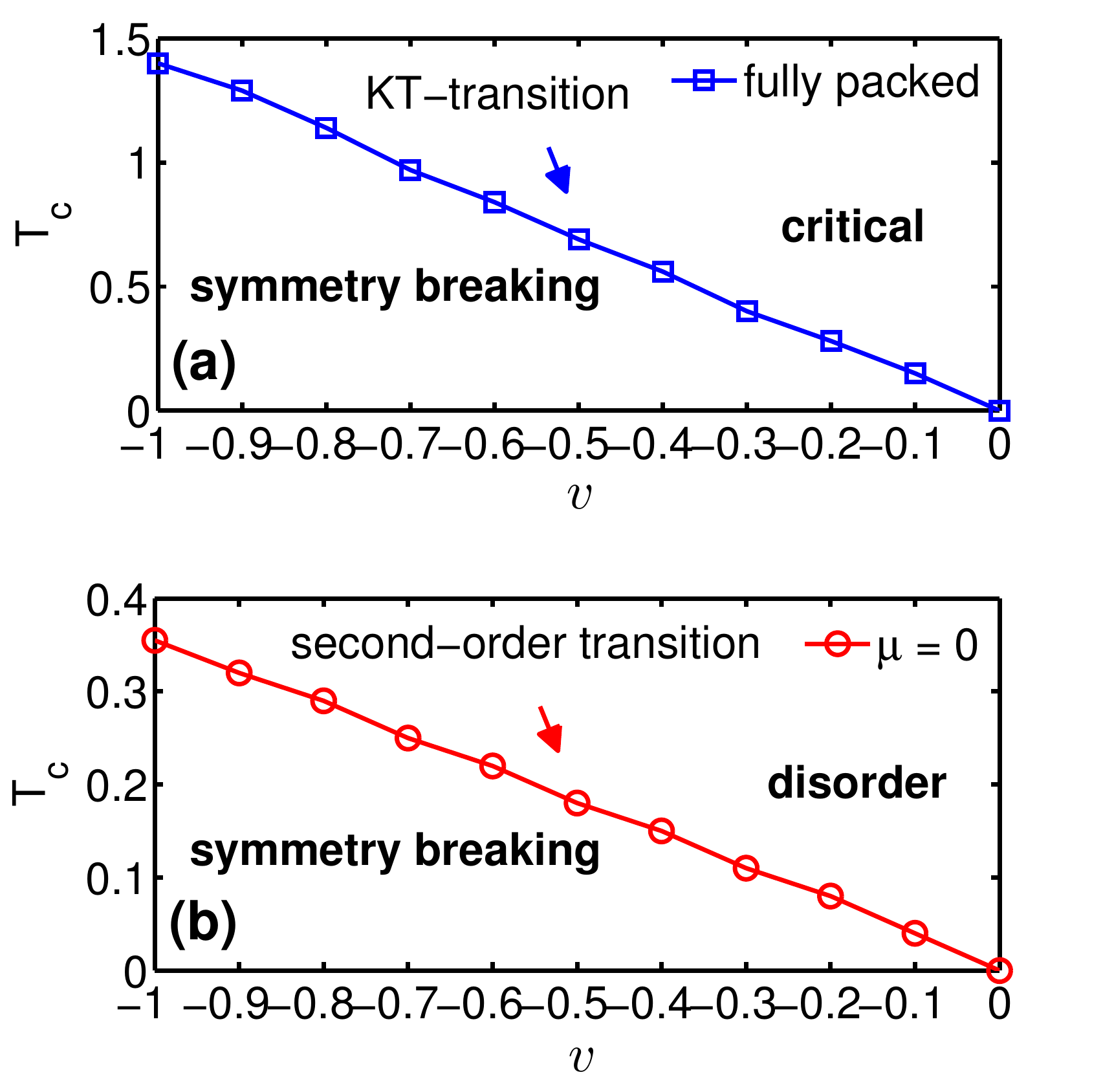}
\caption{(Color online) The phase diagrams of the fully-packed dimer and the monomer-dimer ($\mu=0$) models on the checkerboard lattice. (a) The low-T symmetry breaking phase and the high-T critical phase are separated by a KT phase transition line. (b) The low-T symmetry breaking and the high-T disordered phases are separated by a second-order phase transition line.
\label{phase-v}}
\end{figure}

As a summary of the previous studies of the phase transitions, we show the phase diagrams of the monomer-dimer model in Figs.~\ref{phase-u} and \ref{phase-v}. The $\mu-T$ phase diagram (with fixed $v=-1$) is shown in \Fig{phase-u}. The red vertical line at $\mu = \infty$ is a line consisting of critical points, i.e., a critical line, and the KT-transition point $T_c \approx 1.4$ separates the low-T symmetry breaking phase and the high-T critical phase. When $\mu$ is finite, the phase boundary (blue curved) line represents continuous (second-order) phase transitions, separating the low-T ordered and the high-T disordered non-critical phases. The blue curved line terminates at $\mu_{c}$, which is denoted by a blue star in the inset of \Fig{phase-u}. We estimate, by a polynomial fitting, that $\mu_{c}\approx-0.51$, below which the low-T symmetry breaking phase disappears.

The $\nu-T$ phase diagrams of the fully-packed dimer and the monomer-dimer ($\mu=0$) cases on the checkerboard lattice are shown in \Fig{phase-v}. The phase boundary line of the fully-packed dimer model is a KT phase transition line, which separates the low-T ordered and the high-T critical phases. On the other hand, the phase boundary in the monomer-dimer model with $\mu=0$ is a second-order phase transition line, which separates the low-T ordered and the high-T disordered non-critical phases. Notice that in both cases, the transition temperature $T_c$ vanishes when $\nu=0$, in agreement with the observation that the low-T symmetry breaking phase is induced by the dimer-dimer attractive interactions $\nu$.

\section{conclusion and outlook}
By employing the accurate tensor network method, we have systematically studied the interacting monomer-dimer model on the checkerboard lattice. The specific heat $C_{V}$ and the order parameter $|n_A-n|$ show that KT phase transitions occur in the interacting fully-packed dimer model ($\mu=\infty$), in contrast to the finite-$\mu$ case where second-order phase transitions take place. Collecting the phase transition points, we obtain the $\mu-T$ and $\nu-T$ phase diagrams with fixed $\nu=-1$ and $\mu=0$ or $\infty$, respectively. From the phase diagrams, we find that the attractive interactions $\nu<0$ always induce a symmetry breaking phase at low temperatures, no matter in the fully-packed case ($\mu=\infty$) or the monomer-dimer case ($\mu < \infty$). Previously, people have found similar conclusions for the square-lattice interacting dimer models \cite{Alet-2005}. Here we show that even switching off the interactions on one half of the plaquettes (thus reducing to a checkerboard lattice model), there is still a symmetry breaking dimer crystal phase at a low-T. As a consequence, the dimer crystal does not break the $90$ degree lattice rotational symmetry on the checkerboard lattice.

The efficient tensor network technique enables us to calculate the thermodynamic properties of the monomer-dimer models with a very high precision. For example, the monomer-dimer constant can be determined to the machine precision. The tensor network method also provides novel tools (for example, boundary MPS entanglement entropy) for detecting the phase transitions.

Besides the square and checkerboard lattices, it calls for more investigations of this interacting monomer-dimer models on other lattices, say kagome or star lattice, to explore the dimer-dimer interaction effects in more general situations. The tensor network method is also applicable for investigation of these lattice dimer models and we will discuss them elsewhere.

\section{Acknowledgement}
This work was supported in part by the National Natural Sciences Foundation of China (Grants No. 11274033, and No. 11474015), Major Program of Instrument of the National Natural Sciences Foundation of China (Grant No. 61227902), Sub Project No. XX973 (XX5XX), and the Research Fund for the Doctoral Program of Higher Education of China (Grant No. 20131102130005). W.L. acknowledges the hospitality of the Kavli Institute of Theoretical Physics China where part of this work was performed. W.L. was supported by the DFG through Grant No. SFB-TR12 and Cluster of Excellence NIM.

\end{document}